\documentclass[11pt,english]{article}
\usepackage{jheppub}

\usepackage[T1]{fontenc}
\usepackage[latin9]{inputenc}
\usepackage{amsmath}
\usepackage{esint}
\usepackage{babel}
\usepackage{color}
\usepackage{enumitem}
\usepackage{caption}
\usepackage{subcaption}

\usepackage[numbers,sort&compress]{natbib}
\usepackage{hyperref}
\allowdisplaybreaks
\hypersetup{
colorlinks=true,
linkcolor=blue,
linktocpage=true,
citecolor=violet
}

\usepackage{xargs}
\usepackage[colorinlistoftodos,prependcaption,textsize=tiny]{todonotes}

\def\eps{\epsilon}

\def\res{\mathop{\text{Res}}}

\def\LS{\text{LS}}

\def\PLT{{{\sc PolyLogTools}}}

\newcommand{\munich}{Max-Planck-Institut f\"ur Physik, Werner-Heisenberg-Institut, 
80805 M\"unchen, Germany.}

\begin{document}
\preprint{MPP-2022-131}

\title{One-loop hexagon integral to higher orders in the dimensional regulator}

\author[a]{Johannes M. Henn}
\author[a]{Antonela Matija\v{s}i\'{c}}
\author[a]{and Julian Miczajka}
\affiliation[a]{\munich}
\emailAdd{henn@mpp.mpg.de}
\emailAdd{amatijas@mpp.mpg.de}
\emailAdd{miczajka@mpp.mpg.de}

\abstract{
The state-of-the-art in current two-loop QCD amplitude calculations is at five-particle scattering. Computing two-loop six-particle processes requires knowledge of the corresponding one-loop amplitudes to higher orders
in the dimensional regulator. In this paper we compute analytically the one-loop hexagon integral via differential equations. In particular we identify its function alphabet for general $D$-dimensional external states. We also provide integral representations for all one-loop integrals up to weight four. With this, the one-loop integral basis is ready for two-loop amplitude applications. We also study in detail the difference between the conventional dimensional regularization and the four-dimensional helicity scheme at the level of the master integrals and their function space.
}

\maketitle

\graphicspath{{figs/}}

\section{Introduction}
\label{sec:intro}

There are strong phenomenological and theoretical motivations for studying scattering processes for multi--parton processes. 
On the phenomenology side, matching the expected precision data from the LHC requires perturbative results in quantum field theory.
On the theoretical side, having analytical expressions is crucial for a better understanding of the structure of the amplitudes which may ultimately lead to completely novel approaches for obtaining them.

The current state--of--the--art in multi--parton processes in QCD are at the two--loop five--parton level, for a review see \cite{Zoia:2021zmb}. 
To achieve this, several bottlenecks had to be overcome. For example, finite field methods \cite{Kant:2013vta,vonManteuffel:2014ixa,Peraro:2019svx} helped to tame the necessary computer algebra for relating the Feynman integrals to an integral basis, and likewise for simplifying their (rational, and sometimes algebraic) prefactors.  The necessary two--loop Feynman integrals (with five on--shell legs \cite{Gehrmann:2015bfy,Gehrmann:2018yef,Abreu:2018aqd,Chicherin:2018old}; with four on--shell legs and one off--shell leg \cite{Abreu:2020jxa}) were computed analytically with the help of the canonical differential equations method \cite{Henn:2013pwa}. The analytic answers were further optimized into fast and reliable computer codes capable of evaluating any of the functions in the physical scattering region \cite{Chicherin:2020oor,Chicherin:2021dyp}. 
In consequence, various analytic results for scattering amplitudes have become available, see e.g. \cite{Badger:2017jhb,Badger:2018enw,Badger:2019djh,Abreu:2019odu,Abreu:2021oya,Chawdhry:2021mkw,Abreu:2021asb,Badger:2022ncb}.
Moreover, different groups are already building upon these amplitudes for phenomenological applications, see e.g. \cite{Chawdhry:2021hkp,Czakon:2021mjy,Hartanto:2022qhh}.

In contrast, very little is known at present about two--loop six--particle scattering processes (with the exception of the special all--plus helicity configuration \cite{Dunbar:2016gjb,Dalgleish:2020mof}). 
A first study \cite{Henn:2021cyv} considered the genuine six--particle planar Feynman integrals and found a convenient basis of Feynman integrals that puts the differential equation on the maximal cut into canonical form. 

While the evaluation of the full two-loop Feynman integrals is currently under investigation, it is worthwhile to investigate the corresponding one-loop processes in more detail. Firstly, one key lesson in the computation of five--particle two--loop processes was that important simplifications occur when going from the scattering amplitudes to an appropriate infrared--renormalized finite part (see e.g. \cite{Badger:2019djh}).
For example, the all--plus helicity amplitudes can be written in a stunningly simple, one--line expression. While other helicity amplitudes are more complicated, simplifications both at the level of the transcendental functions, and at the level of the coefficients, were observed. 
While it would be desirable to compute this finite part directly from infrared--finite loop integrals, in the present conventional setup it is obtained as a difference of amplitudes at different loop orders, with the lower--loop amplitudes being multiplied by infrared subtraction terms. In dimensional regularization this implies that the one--loop amplitudes need to be known to higher orders in the dimensional regulator.
Secondly, when computing two--loop cross sections, one--loop amplitudes with additional particles are needed for integration in phase--space integrals. This means that they are needed to an appropriate order in the dimensional regulator, and that particularly fast and reliable numerical representations are required.
However, the analytic results in the literature are typically available only up to the finite part, see e.g. \cite{Dunbar:2009uk,Laurentis:2019bjh}.

Furthermore, we can use two different schemes to perform the computation. Conventional dimensional regularization, where both the loop integration variables, as well as the external variables, are taken to be in D dimensions and four--dimensional helicity scheme, where external states are taken to be in four dimensions. This implies a Gram determinant condition between the kinematic variables and allows for an algebraic reduction of the hexagon integrals in terms of pentagon integrals. The scheme choice is expected to have important practical consequences for the two--loop calculation. For example, it is known that $(D-4)$--dimensional, scheme--dependent, parts of loop integrands are relevant for finding a uniform weight integral basis \cite{Chicherin:2018old,Henn:2021cyv}. It would be very interesting to understand whether there is a natural scheme where the integrands take a simple form, as is the case in four dimensions.

Therefore, we compute the hexagon integrals (together with all integrals in subsectors) in dimensional regularization. In this way we obtain a basis of transcendental functions and  one--fold integral representations that can readily be used for one--loop amplitudes.

We prefer to perform more general calculation which is valid in arbitrary six--particle kinematics for several reasons. 
On the one hand, the full $D$--dimensional result offers more flexibility, as one can vary independently the nine kinematic variables of the process, without regard to the complicated Gram determinant constraint surface. This can be advantageous for several investigations, such as the study of analyticity and Steinmann relations, when taking limits or performing analytic continuation. Treating the internal and external variables on the same footing is also essential if one wishes to study the consequences of conformal symmetry of on--shell processes at a conformal fixed point \cite{Braun:2003rp,Chicherin:2017bxc,Chicherin:2018ubl,Chicherin:2022gky}.
On the other hand, our more general setup allows us to study precisely the difference between the two different schemes. At the level of the results, we will be able to understand at what order the difference start, and what form it takes.

This paper is organized as follows. In section \ref{sec:kinematics} we define our notation and introduce the six--particle on--shell kinematics.
In section \ref{sec:oneloopintegralsandalphabet} we present the one-loop integral basis and the relevant function space.
In section \ref{sec:DE} we present details of our differential equations calculation, explaining in particular how the boundary constants are obtained from simple physical requirements.
We use the analytic results to obtain one--fold integral representations that can readily be evaluated numerically within the Euclidean region. We validate this by performing numerical checks at certain reference points. We analyze in detail the result for the hexagon integral and the alphabet letters that occur in it.
Finally, we study the limit of four-dimensional external states at the level of the differential equations. We conclude and present an outlook in section \ref{sec:outlook}.

This submission includes a list of ancillary files which contain selected results in machine--readable form. We provide the following files:
\begin{itemize}
    \item \verb|PDEMatrix.txt| containing the partial differential equation matrix for the one--loop hexagon family expressed in terms of the alphabet letters,
    \item \verb|HexagonAlphabet.txt| containing the list of alphabet letters,
    \item \verb|HexagonSqrts.txt| containing the substitution rules for the square roots appearing in the alphabet,
    \item \verb|W2Functions.m| containing the basis functions up to weight two and expressions for the master integrals in terms of basis functions,
    \item \verb|BoundaryValue.txt| containing the vector of boundary values up to transcendental weight four,
    \item \verb|NumericalEvaluation.wl| with a proof--of--concept implementation of the one--fold integral representation for numerical evaluation.
\end{itemize}

\section{Six--particle kinematics}
\label{sec:kinematics}

In this paper, we are interested in all integrals from the massless one--loop hexagon family,~i.e.~integrals of the form
\begin{equation}
    I^{(D_0)}(a_1,...,a_6) = e^{\eps \gamma_\text{E}}\int \frac{\text{d}^{D_0-2\epsilon}l}{i\pi^\frac{D_0}{2}} \frac{1}{\prod_{j=1}^6 (l+\sum_{k=1}^{j-1}p_k)^{2a_j}},
\end{equation}
with integer propagator powers $a_i$ and the Euler--Mascheroni constant $\gamma_E$. $D_0$ denotes the number of spacetime dimensions of the integration variables. In principle, this massless six--particle Feynman integral depends on six momenta $p_i\in\mathbb{R}^{D_{\text{ext}}}$ which satisfy $p_i^2=0$ and momentum conservation implies
\begin{align}
    \sum_{i=1}^6 p_i = 0\label{eq:6ptVariables}
\end{align}
if all momenta are taken to be incoming.  However, due to Poincar\'{e} symmetry, the kinematic dependence simplifies and an appropriate set of variables in integer dimensions $D_{\text{ext}}>4$ are the nine independent Mandelstam invariants
\begin{align}
    \vec v = \{s_{12}, s_{23},s_{34},s_{45},s_{56},s_{61}, s_{123},s_{234},s_{345}\}
\end{align}
with
\begin{align}
    s_{ij} = (p_i + p_j)^2, \qquad s_{ijk} = (p_i+p_j+p_k)^2.
\end{align}
We also introduce the cyclic permutation operator $T$ that shifts the external legs by one site,
\begin{equation}
    T(p_i)=p_{i+1}, \quad i=1,\ldots,6
    \label{eq:CyclicShift} 
\end{equation}
and it acts on the variables according to
\begin{equation}
    T \vec{v} = \{s_{23}, s_{34}, s_{45}, s_{56}, s_{61}, s_{12}, s_{234}, s_{345}, s_{123}\}.
\end{equation}

For physical momentum configurations describing $2\rightarrow 4$ or $3 \rightarrow 3$ scattering processes, the Mandelstam invariants take definite signs depending on which particles are incoming and outgoing respectively. Additionally, there are certain Gram determinant constraints that are required to hold in all physical regions. The Gram determinants $G$ are defined as
\begin{equation}
    G(q_1,...,q_n;u_1,...,u_n)=\det(2q_i\cdot u_j), \quad 1 \le i,j \le n,
\end{equation}
with 
\begin{equation}
G(p_1,...,p_n) = G(p_1,...,p_n;p_1,...,p_n).
\label{eq:GramDet}
\end{equation}
We also introduce the notation for the modified Gram determinant of an integral with external momenta $p_j$,
\begin{equation}
    G^\star(q_1, \dots, q_{n-1};u_1,\dots, u_{n})=\left.G(l,q_1,...,q_{n-1};u_1,\dots, u_{n})\right|_{l\cdot p_j = -\sum_{k<j} p_k\cdot p_j-\frac{1}{2} p_j^2}.
    \label{eq:graml}
\end{equation}
Naturally, for the hexagon integral we have $p_j^2 = 0$, whereas some of the subsectors of pentagons, boxes, triangles and bubbles will have composite external momenta which are not necessarily on-shell.

Defining further 
\begin{equation}
    \bar{G}_n = (-1)^{n-1}G(p_1,...,p_n), \quad 1\le n \le 6
\end{equation}
the Gram determinant constraints take the form~\cite{Byckling:1971vca}
\begin{equation}
    \bar{G}_1 = 0, \quad \bar{G}_2 > 0, \quad \bar{G}_3 > 0, \quad \dots, \quad \bar{G}_{D_0} > 0
    \label{eq:GramPositivityConstraints}
\end{equation}
and $\bar{G}_n = 0$ for all $n>D_0$, since there are only $D_0$ independent momenta in $D_0$ dimensions.
However, in parameter integral representations, it is perfectly reasonable to analytically continue the Feynman integrals beyond the physical regions where these constraints are satisfied. In this paper, we will focus on the so called Euclidean region, in which all Mandelstam invariants are negative,
\begin{equation} 
s_{ij}<0, \qquad s_{ijk} <0.
\end{equation}
Most of our calculations will be performed in this unphysical region, with the understanding that results acquired there can be transported to physical regions by analytical continuation. The main advantage of working in the Euclidean region is that dimensionally regularised Feynman integrals are real-valued all throughout it and only diverge on the boundaries where some of the Mandelstam invariants vanish. 

For completeness, let us note the signs that the Mandelstam invariants acquire in selected physical scattering regions \cite{Byckling:1971vca}. 
\paragraph{12$\rightarrow$3456.}In the $2\rightarrow4$ scattering region with particles $1$ and $2$ incoming, in addition to the Gram determinant constraints \eqref{eq:GramPositivityConstraints}, the constraints on the Mandelstam invariants are
\begin{align}
    s_{12}, s_{34}, s_{35},s_{36},s_{45},s_{46},s_{56} &>0, \notag \\
    s_{13}, s_{14}, s_{15}, s_{16}, s_{23},s_{24}, s_{25}, s_{26}&<0.
    \end{align}
    Note that the constraints that go beyond the set of variables introduced in \eqref{eq:6ptVariables} can be reexpressed in terms of the $v_i$. For example, 
    \begin{equation}
        s_{35} = s_{345} - s_{34} - s_{45}.  
    \end{equation}
\paragraph{123$\rightarrow$ 456.}  In the $3\rightarrow 3$ scattering region with particles $1, 2$ and $3$ incoming, in addition to the Gram determinant constraints \eqref{eq:GramPositivityConstraints}, the Mandelstam invariants satisfy
\begin{align}
    s_{12}, s_{13}, s_{23}, s_{45}, s_{46}, s_{56} &> 0, \notag\\
    s_{14}, s_{15}, s_{16}, s_{24}, s_{25}, s_{26}, s_{34}, s_{35}, s_{36} &< 0.
\end{align}
\paragraph{Gram determinant constraints in $D_{\text{ext}}=4$ dimensions.} Quite importantly, the nine variables $v_i$ are only independent in an arbitrary number of dimensions $D_{\text{ext}}>4$. In $D_{\text{ext}}=4$ dimensions, both $\bar{G}_5$ and $\bar{G}_6$ are required to vanish. While $\bar{G}_6$ vanishes trivially by momentum conservation, the vanishing of $\bar{G}_5=0$ imposes a non-trivial constraint on the Mandelstam variables, effectively reducing the number of independent variables from nine to eight. The constraint is a degree--five polynomial in the variables $v_i$ that is quadratic in any of the $v_i$. It is also invariant under arbitrary permutations of the external momenta.

A parametrisation of the external degrees of freedom that hardwires this constraint, as well as momentum conservation and on--shellness can be provided by choosing a momentum twistor configuration~\cite{Hodges:2009hk}. In this paper, we employ the particular parametrisation~\cite{Henn:2021cyv,Badger:2013gxa}
\begin{align}
    v_1&=x_1, \notag \\
    v_2&= x_1 x_5, \notag \\
    v_3&= \dfrac{x_1 \left[x_5-x_2x_3x_6+x_3x_5\left(1+x_2-x_2x_7\right)\right]}{x_2}, \notag \\
    v_4&=x_1\left[x_5-x_5x_7-\left(1+x_3\right)x_4\left(x_5\left(-1+x_7\right)+x_8\right) + \dfrac{x_2x_3x_4\left(x_5\left(-1+x_7\right)+x_6x_8\right)}{x_5}\right], \notag \\
    v_5&= \dfrac{x_1x_3\left[\left(x_2-x_5\right)x_5\left(-1+x_7\right)+\left(-x_5+x_2x_6\right)x_8\right]}{x_5}, \notag \\
    v_6&= \dfrac{x_1x_2x_3x_4\left[x_5\left(-x_6+x_7\right)+x_6x_8\right]}{x_5}, \notag \\
    v_7 &= x_1x_8, \notag \\
    v_8&= x_1x_3\left(x_2x_6-x_5x_7\right), \notag \\
    v_9&=x_1 \left\{x_6+x_4\left\{-1+x_6+x_3\left[-1+x_6+x_2\left(-1+x_7+ \dfrac{x_6x_8}{x_5}\right)\right]\right\}\right\}
    \label{eq:MomTwistRepl}
\end{align}
where $x_i \in \mathbb{R}$ are unconstrained variables.

\section{One--loop integrals and function alphabet}
\label{sec:oneloopintegralsandalphabet}
In this section, we construct a canonical basis of master integrals for the hexagon family by using dimension--shift identities and a Baikov analysis to normalise the integrals by their leading singularities. We then describe the alphabet of letters required to express the canonical differential equation for this basis in terms of dlog forms.
\subsection{Master integrals}
Using integration-by-parts (IBP) identities~\cite{Laporta:2000dsw}, all of the integrals in the hexagon family can be reduced to a basis of 33 master integrals. We use a combination of {\verb|FIRE|}~\cite{Smirnov:2008iw} and {\verb|LiteRed|}~\cite{Lee:2012cn} to perform the reductions automatically. A convenient basis choice $I_j$ is spanned by\footnote{We employ the permutation operator defined in~\eqref{eq:CyclicShift}.}:
\begin{figure}
    \centering
    \begin{subfigure}[b]{0.2\textwidth}
    \centering
    \includegraphics[width=\textwidth]{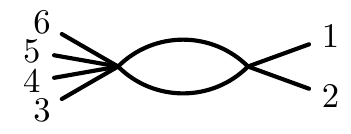}
    \caption{$I_1$}
    \label{fig:I1}
    \end{subfigure}
    \hfill
    \begin{subfigure}[b]{0.2\textwidth}
    \centering
    \includegraphics[width=\textwidth]{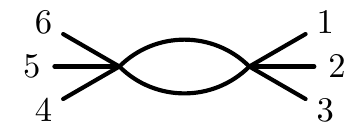}
    \caption{$I_7$}
    \label{fig:I7}
    \end{subfigure}
    \hfill
    \begin{subfigure}[b]{0.15\textwidth}
    \centering
    \includegraphics[width=\textwidth]{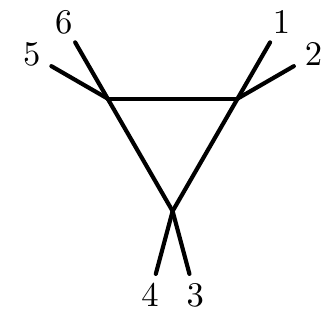}
    \caption{$I_{10}$}
    \label{fig:I10}
    \end{subfigure}
    \hfill
    \begin{subfigure}[b]{0.15\textwidth}
    \centering
    \includegraphics[width=\textwidth]{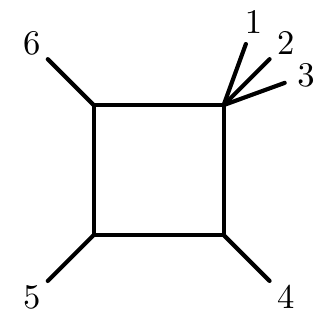}
    \caption{$I_{13}$}
    \label{fig:I13}
    \end{subfigure}

    \begin{subfigure}[b]{0.18\textwidth}
    \centering
    \includegraphics[width=\textwidth]{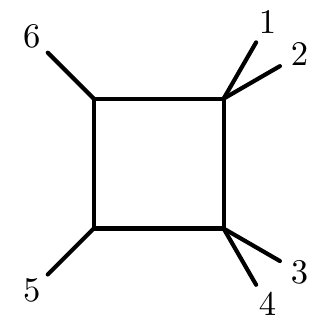}
    \caption{$I_{19}$}
    \label{fig:my_label}
    \end{subfigure}
    \hfill
    \begin{subfigure}[b]{0.18\textwidth}
    \centering
    \includegraphics[width=\textwidth]{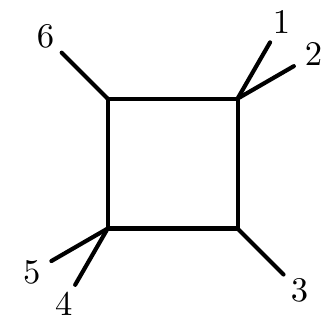}
    \caption{$I_{25}$}
    \label{fig:I19}
    \end{subfigure}
    \hfill
    \begin{subfigure}[b]{0.18\textwidth}
    \centering
    \includegraphics[width=\textwidth]{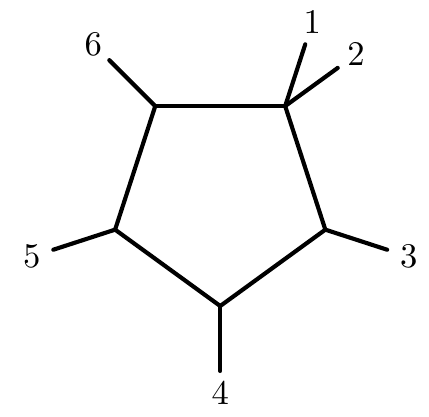}
    \caption{$I_{28}$}
    \label{fig:I28}
    \end{subfigure}
    \hfill
    \begin{subfigure}[b]{0.18\textwidth}
    \centering
    \includegraphics[width=\textwidth]{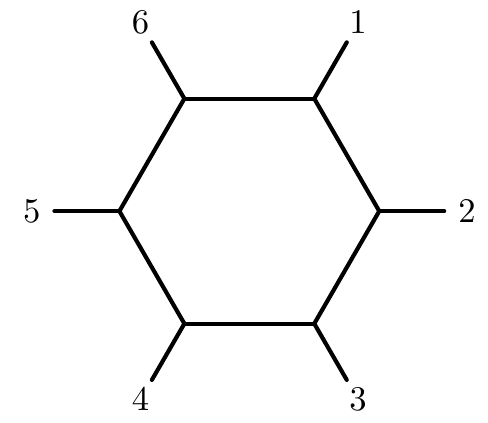}
    \caption{$I_{33}$}
    \label{fig:I33}
    \end{subfigure}
    \caption{Graphical representations of the integrals in the IBP basis.}
    \label{fig:foobar}
\end{figure}

\begin{itemize}
    \item the six cyclic permutations of the massive bubble integral \begin{equation}I_{i}=T^{i-1} I^{(4)}(1,0,1,0,0,0),\quad i=1,...,6,\end{equation} 
    \item the three cyclic permutations of the massive bubble integral \begin{equation}I_{6+i}=T^{i-1} I^{(4)}(1,0,0,1,0,0),\quad i=1,...,3,\end{equation} 
    \item the two cyclic permutations of the three--mass triangle integral \begin{equation}I_{9+i}=T^{i-1} I^{(4)}(1,0,1,0,1,0),\quad i=1,2,\end{equation} 
    \item the six cyclic permutations of the one--mass box integral \begin{equation}I_{11+i}= T^{i-1}I^{(4)}(0,0,1,1,1,1),\quad i=1,...,6,\end{equation} 
    \item the six cyclic permutations of the two--mass--hard box integral \begin{equation}I_{17+i}=T^{i-1} I^{(4)}(0,1,0,1,1,1),\quad i=1,...,6,\end{equation} 
    \item the three cyclic permutations of the two--mass--easy box integral \begin{equation}I_{23+i}=T^{i-1} I^{(4)}(0,1,1,0,1,1), \quad i=1,2,3,\end{equation} 
    \item the six cyclic permutations of the one--mass pentagon integral \begin{equation}I_{26+i}=T^{i-1} I^{(4)}(0,1,1,1,1,1), \quad i=1,...,6,\end{equation} 
    \item the hexagon integral \begin{equation}I_{33}=I^{(4)}(1,1,1,1,1,1).\end{equation}  
\end{itemize}
However, the integrals in this basis do not have uniform transcendentality (UT) and therefore do not satisfy a differential equation in canonical form. 
\subsection{Constructing a UT basis}
To determine a UT basis, we use the observation that $n$--point and $(n-1)$--point integrals in $D=n-2\epsilon$ dimensions have uniform transcendentality if they are normalised by their leading singularities in $D=n$ dimensions~\cite{Spradlin:2011wp}. Hence, for the construction of a UT basis we prefer to include integrals in $D=2-2\epsilon$ and $D=6-2\epsilon$ dimensions.
In fact, if $D_0$ differs from four by integer multiples of two, these integrals are related to the $D_0=4$ integrals by dimension shift identities~\cite{Tarasov:1996br,Lee:2009dh}. Explicitly, this allows us to express bubble integrals in two dimensions through bubble integrals in four dimensions as
\begin{equation}
    I^{(2)}(1,0,1,0,0,0) = \frac{2(1-2\epsilon)}{s_{12}} I^{(4)}(1,0,1,0,0,0),
\end{equation}
as well as pentagon integrals in six dimensions through box and pentagon integrals in four dimensions 
\begin{equation}
    I^{(6)}(0,1,1,1,1,1) = \frac{1}{2\epsilon G(p_2,p_3,p_4,p_5)} \vec{c}_P \cdot \vec{I}_P,
\end{equation}
where
\begin{equation}
    \vec{I}_P = \begin{pmatrix}I_{12} & I_{17} & I_{18} &  I_{22}& I_{24} & I_{27}\end{pmatrix}
\end{equation}
and
\begin{align}
c_{P1}&= v_3 v_4 (v_8 v_9 - v_3 v_6 + v_2 (v_9-v_3)+v_4(v_3-v_8)),\notag\\
c_{P2}&= v_2 v_3 (v_2 (v_3 - v_9)-v_3 (v_6+v_4) + v_8 (v_9+v_4)),\notag\\
c_{P3}&= v_4 (-2 v_2 v_6 v_3 + v_2 v_8 (v_3 + v_9) + 
  v_8 (v_6 v_3 - v_3 v_4 + v_8 (-v_9 + v_4))),\notag\\
c_{P4}&=v_2 (-2 v_6 v_3 v_4 + v_3 v_9 (-v_2 + v_6 + v_4) + 
   v_9 (v_2 v_9 + v_8 (-v_9 + v_4))),\notag\\
c_{P5}&=(v_6 v_3 - v_8 v_9) (v_6 v_3 - v_8 v_9 + v_2 (-v_3 + v_9) + 
   v_8 v_4 - v_3 v_4), \notag\\
c_{P6}&= 2 v_2 v_3 v_4 (v_6 v_3 - v_8 v_9).
\end{align}
Finally, the six--dimensional hexagon integral can be decomposed into four--dimensional pentagon and hexagon integrals according to 
\begin{equation}
    I^{(6)}(1,1,1,1,1,1) = \frac{1}{2(1+2\epsilon)G(p_1,p_2,p_3,p_4,p_5)}\vec{c}_H \cdot \vec{I}_H,
\end{equation}
where 
\begin{equation}
    \vec{I}_H = \begin{pmatrix} I_{27} & I_{28} & I_{29} & I_{30} & I_{31} & I_{32} & I_{33} \end{pmatrix}
\end{equation}
and\footnote{Note that the Gram determinant on the maximal cut should be evaluated before acting with the cyclic permutation operator. }
\begin{align}
c_{Hi} &= - T^{i} G^\star(p_1,\dots,p_4;p_5,p_1,\ldots,p_4), \quad i=1,\dots,6\notag\\
c_{H7} &=  -G^\star(p_1, \ldots, p_5).
\label{eq:DimShiftHex}
\end{align}
With these dimension--shift identities, we can set up an intermediate basis, where the bubbles are two--dimensional, the triangles and boxes four--dimensional and the pentagons and hexagons six--dimensional integrals,~i.e.~
\begin{align}
    \bar{I}_{i} &= \hat{d}^{-2}I_i, & i&=1,\dots,9, \notag\\
    \bar{I}_{i} &= I_i, & i&=11,\dots,26, \notag\\
    \bar{I}_{i} &= \hat{d}^2 I_i, & i&=27,\dots,33
\end{align}
with the dimension--shift operator $\hat{d}$. This basis is still not of universal transcendentality. 
\paragraph{Leading singularities from the Baikov representation.}\label{subsec:leadingsingularities}
To transform the above basis of master integrals into a basis where every single element is a pure function of uniform transcendentality, we normalise the integrals by their leading singularities. Calculating the leading singularities of one--loop integrals is most straightforward in the Baikov representation~\cite{Baikov:1996iu, Henn:2021cyv, Dlapa:2021qsl}. 
For an $L$--loop integral with $E$ external points and $n$ propagators $D_i$
\begin{equation}
    I(a_1,\dots,a_n) = \int  \frac{\text{d}^D l}{i \pi^{D/2}} \frac{1}{D_1^{a_1}\dots D_n^{a_n}},
\end{equation}
the Baikov representation is reached by the change of variables $z_i = D_i$ and reads
\begin{equation}
    I(a_1,...,a_n) = C(E) \tilde{G}^{\frac{E-D}{2}} \int \frac{\text{d}z_1 \dots \text{d}z_n}{z_1^{a_1} \dots z_n^{a_n}} S^{\frac{D-E-1}{2}},
\end{equation}
where $\tilde{G}$ and $S$ are Gram determinants given by
\begin{align}
    \tilde{G} &= G(p_1,\dots,p_{E-1})\notag\\
    S &= \left.G(l,p_1,...,p_{E-1})\right|_{l\cdot p_j = -\sum_{k<j} p_k\cdot p_j-\frac{1}{2} (z_{j}-z_{j+1}+p_j^2)}.
\end{align}
The leading singularity can then be extracted straightforwardly by taking the residue at~$z_i = 0$. $C(E)$ is a kinematic-independent constant that has no impact on the UT property of an integral and can be omitted. Hence, the leading singularity of an integral takes the form
\begin{equation}
    \LS(I(a_1,\dots,a_n)) = G^{\frac{E-D}{2}}\res_{z_i=0} \frac{S^{\frac{D-E-1}{2}}}{{z_1^{a_1} \dots z_n^{a_n}}}.
\end{equation}
The expression becomes even simpler if all propagator powers $a_j=1$:
\begin{equation}
    \LS(I^{(D)}(1,\dots,1)) = G^{\frac{E-D}{2}} \lim_{z_i\rightarrow 0} S^{\frac{D-E-1}{2}}.
\end{equation}
Using the definition of $G^\star$ in eq.~\eqref{eq:graml}, for the one--loop $n$--point integrals in $n$ or $n+1$ dimensions we can further rewrite this as
\begin{align}
    \LS(I^{(D)}(\underbrace{1,\dots,1}_{n\text{ entries}}))=\left\{\begin{array}{l} \sqrt{G^\star( p_1,....,p_{n-1})}^{-1}, \qquad \text{if } D=n,\\
    \sqrt{G(p_1,...,p_{n-1})}^{-1}, \qquad \text{if } D=n+1.
    \end{array}\right.
\end{align}
With this formula, we can straightforwardly calculate the leading singularities for our basis of integrals, namely\footnote{Note that due to \eqref{eq:GramPositivityConstraints}, the leading singularities of the three--mass triangle and of the one--mass pentagon integrals are purely imaginary in physical regions.}
\begin{align}
    \LS(I^{(2)}(1,0,1,0,0,0)) &= v_{1}^{-1}, \notag\\ \text{LS}(I^{(2)}(1,0,0,1,0,0)) &= v_{7}^{-1},\notag\\
    \LS(I^{(4)}(1,0,1,0,1,0)) &= \sqrt{\Delta_3(v_{1},v_{3},v_{5})}^{-1},\notag\\ \LS(I^{(4)}(0,0,1,1,1,1)) &= (v_{3}v_{4})^{-1},\notag\\
    \LS(I^{(4)}(0,1,0,1,1,1)) &= (v_{8}v_{4})^{-1}, \notag\\
    \LS(I^{(4)}(0,1,1,0,1,1)) &= (v_{8}v_{9}-v_{3}v_{6})^{-1}\notag\\
    \LS(I^{(6)}(0,1,1,1,1,1)) &= \sqrt{\Delta_5(v_{2},v_{3},v_{4},v_{8},v_{9},v_{6})}^{-1}\notag\\
    \LS(I^{(6)}(1,1,1,1,1,1)) &= \sqrt{\Delta_6}^{-1},
\end{align}
and the permutations thereof, where we defined
\begin{align}
\Delta_3(v_{1},v_{3},v_{5})=\lambda(v_{1},v_{3},v_{5}), \qquad\qquad \qquad\quad\;&\notag\\
\Delta_5(v_{2},v_{3},v_{4},v_{8},v_{9},v_{6}) = (\omega_5(v_{2},v_{3},v_{4},v_{8},v_{9})-v_{3}v_{6})^2&\notag\\
\qquad \qquad \qquad \quad\qquad \quad \ -4 v_{2} v_{3} v_{4} (  v_{3} +v_{6}-v_{8}- v_{9} ),& \notag\\
\Delta_6 = ((1+T+T^2)(v_{1}v_{8}v_{4})-v_{7}v_{8}v_{9})^2 -4v_{1}v_{2}v_{3}v_{4}v_{5}v_{6},&
\end{align}
and
\begin{gather}
    \lambda(v_{1},v_{3},v_{5})=v_{1}^2+v_{3}^2+v_{5}^2-2v_{1}v_{3}-2v_{1}v_{5}-2v_{3}v_{5},\notag\\
    \omega_5(v_{2},v_{3},v_{4},v_{8},v_{9},v_{6}) =v_{8}v_{9}+v_{2}(v_{3}-v_{9})+v_{4}(v_{3}-v_{8}).
\end{gather}
Interestingly, $\Delta_5$ takes a strikingly simple form in terms of the K\"all\'en function $\lambda$, e.g.~
\begin{equation}
    \Delta_5(v_1,v_2,v_3,v_7,v_8,v_5) = \lambda(v_1 v_3,(v_7-v_1-v_2)(v_8-v_2-v_3), v_2 (v_5-v_7-v_8+v_2)).
\end{equation}
Finally, having started with the four--dimensional basis $I$, after using dimension--shift identities to construct the basis $\bar{I}$ we can further transform it into a basis of uniform transcendentality $\tilde{I}$, by normalising it according to
\begin{align}
    \tilde{I}_i = \frac{\bar{I}_i}{\LS(\bar{I}_i)}\times \left\{\begin{array}{ll}\epsilon, &\quad i=1,\dots,9,\\\epsilon^2, &\quad i=10,\dots,26,\\\epsilon^3, &\quad i=27,\dots,33.\end{array}\right.
    \label{eq:UTBasis}
\end{align}
Here, the explicit powers of $\epsilon$ make sure that not only is every single integral of uniform transcendentality but that all the different integrals also have vanishing transcendent weight, c.f.~\cite{Hannesdottir:2021kpd}.\footnote{Recall that $\eps$ has transcendental weight $-1$.} Since the finite part of the pentagon and hexagon integrals in four dimensions are transcendental functions of weight three, their epsilon expansion in arbitrary dimensions starts at order $\epsilon^3$ and the first correction to the four-dimensional result is of order $\epsilon^4$.
\subsection{Function space up to weight two}\label{subsec:fweighttwo}
Hence, our goal is to obtain the one--loop hexagon integral up to fourth order in the dimensional regulator $\epsilon$. Up to weight two, our basis integrals can be expressed in terms of a basis of special functions, while the weight--three and weight--four parts can be expressed as a one--fold integrals over the weight--two functions (see section \ref{subsec:intrepw4}). 

Up to weight two, only the bubbles, triangles and boxes contribute, hence the weight--two functions are already known, see e.g.~Ref.~\cite{Gehrmann:2018yef,Chicherin:2020oor,Chicherin:2021dyp}. We denote the functions with $f_i^{(j)}$ where $i$ is a label and $j$ is the transcendental weight.  

The basis of weight--one functions consists of
\begin{gather}
   f^{(1)}_1 =  \log (-v_1), \quad f_{i+1}^{(1)}=T^if_1^{(1)}, \quad i=1,\ldots,5\\
   f_7^{(1)} = \log (-v_7),  \quad f_{i+7}^{(1)}=T^if_7^{(1)}, \quad i=1,\ldots,2
   \label{eq:w1}
\end{gather}
where the minus sign is used in order for the functions to be well defined in the Euclidean region.

At weight two, we construct the initial basis as
\begin{equation}
    \{f^{(2)}, (f^{(1)}f^{(1)}), \zeta_2 \},
\end{equation}
where $f^{(2)}$ are weight--two functions and $(f^{(1)}f^{(1)})$ are products of two weight--one functions given in (\ref{eq:w1}). Genuine weight--two functions that can appear are dilogarithms $\mathrm{Li}_2$. We chose the arguments of these functions in such a way that they are well defined in the Euclidean region.

The basis of weight--two functions consists of 12 dilogarithmic functions depending on two kinematic variables
\begin{gather}
    f_1^{(2)}=\mathrm{Li}_2 \left(1-\dfrac{v_1}{v_7}\right),\quad f_{i+1}^{(2)}=T^if_1^{(2)}, \quad i=1,\ldots,5, \\
    f_7^{(2)}=\mathrm{Li}_2 \left(1-\dfrac{v_2}{v_7}\right),\quad f_{i+7}^{(2)}=T^if_7^{(2)}, \quad i=1,\ldots,5.
\end{gather}
These functions appear as part of the expression for one--mass and two--mass--hard box integrals. \\
Next, we have three dilogarithms depending on four kinematic variables which are part of the expression for two--mass--easy box integrals
\begin{equation}
    f_{13}^{(2)}=\mathrm{Li}_2 \left(1-\dfrac{v_1v_4}{v_7v_9}\right),\quad f_{i+13}^{(2)}=T^if_{13}^{(2)}, \quad i=1,\ldots,2.
\end{equation}
For example, one of the two--mass--easy box integrals is given explicitly by the following linear combination of basis functions:
\begin{align}
    \tilde{I}_{26}=&\ 2\bigg[\eps \left(f^{(1)}_2 + f^{(1)}_5-f^{(1)}_7-f^{(1)}_8\right)  \notag \\
     &\ +\eps^2\Big(-\frac{1}{2}(f^{(1)}_2)^2 -\frac{1}{2}(f^{(1)}_5)^2 + f^{(1)}_7f^{(1)}_8-f^{(2)}_7-f^{(2)}_{10}- f^{(2)}_2-f^{(2)}_5+f^{(2)}_{14}\Big)\bigg].
\end{align}
Finally, the remaining two weight--two functions are 
\begin{equation}
    f_{16}^{(2)}=\mathrm{Tri} (v_1,v_3,v_5),\quad f_{17}^{(2)}=Tf_{17}^{(2)},
\end{equation}
where $\mathrm{Tri} (a,b,c)$ is the three--mass triangle function defined as
\begin{align}
    \mathrm{Tri} (a,b,c) &= -\text{Li}_2\left(-\frac{2 b}{a-b-c-\sqrt{\Delta_3(a,b,c)}}\right) 
    -\text{Li}_2\left(-\frac{2 c}{a-b-c-\sqrt{\Delta_3(a,b,c)}}\right)\notag\\
    &\ -\frac{\pi ^2}{6}
     -\frac{1}{2} \log \left(\frac{c}{b}\right) \log \left(\frac{a-b+c-\sqrt{\Delta_3(a,b,c)}}{a+b-c-\sqrt{\Delta_3(a,b,c)}}\right)\notag\\
     &\ -\frac{1}{2} \log \left(\frac{2 b}{-\sqrt{\Delta_3(a,b,c)}+a-b-c}\right) \log \left(\frac{2 c}{-\sqrt{\Delta_3(a,b,c)}+a-b-c}\right),
    \label{eq:Tri}
\end{align}
and it corresponds to the Bloch--Wigner dilogarithm~\cite{Zagier:2007knq}.
The triangle function (\ref{eq:Tri}) appears as the weight--two expression for the two three--mass triangle integrals in our function basis. These integrals are normalized by the square root of K\"all\'en function, denoted by $\sqrt{\Delta_3}$, that can either be real or imaginary in the Euclidean region. It is convenient to choose a part of the Euclidean region where this square root is imaginary since the three--mass triangle integrals are single--valued there.

In total we have 17 weight--two functions in addition to the products of weight--one functions which we do not list separately. The full basis of weight--two functions as well as the expression for our canonical basis of integrals in terms of these functions is given in the ancillary file \verb|W2Functions.m|.

\subsection{Canonical differential equation}
The basis of UT master integrals \eqref{eq:UTBasis} satisfies a differential equation in canonical form 
\begin{equation}
    \text{d}\tilde I = \eps A \tilde I,
    \label{eq:CanonicalPDE}
\end{equation}
where d is a total differential with respect to Mandelstam invariants $v_i$
\begin{equation}
    \text{d}=\sum_{i=1}^9\text{d}v_i \dfrac{\partial}{\partial_{v_i}}
\end{equation}
and $A$ is a matrix of logarithmic forms
\begin{equation}
    A_{jk}=\sum_{a} c_{jk}^a \text{d}\log(W_a).
    \label{eq:PDEMatrix}
\end{equation}
The matrices $c_{jk}^a$ appearing in \eqref{eq:PDEMatrix} are constants (i.e.~$c_{ij}^k \in \mathbb{Q}$) while the letters $W_a$ are algebraic functions of the kinematic variables taken from a list that we call the hexagon alphabet~$\mathbb{A}$. In particular, the overall factor of $\eps$ is the sole dependence on the dimensional regularisation parameter. 

If the matrix $A$ is known, derivatives of the function basis with respect to any of the variables $v_i$ in \eqref{eq:6ptVariables} are calculated via
\begin{equation}
    \frac{\partial \tilde I}{\partial v_i} = \epsilon A_i \tilde{I}
\end{equation}
with
\begin{equation}
    (A_i)_{jk} = \sum_{a} c_{jk}^a \frac{\partial \log W_a}{\partial v_i}.
    \label{eq:MandelstamPDEs}
\end{equation}
In fact, we use this knowledge to construct the differential equation. It is straightforward to rewrite the derivatives with respect to the Mandelstam invariants in \eqref{eq:MandelstamPDEs} in terms of derivatives with respect to momenta $p_i$ using the chain rule. Practically, this can be done by starting with an ansatz \cite{Henn:2014qga}, e.g.~
\begin{equation}
\frac{\partial}{\partial s_{12}} = \sum_{j=1}^5(\beta_j p_j)\cdot \frac{\partial}{\partial p_1}
\end{equation}
and imposing compatibility with the on--shell constraints $p_i^2=0$ and $(\sum_{i=1}^5 p_i)^2$ as well as the defining relations 
\begin{equation}
    \frac{\partial}{\partial v_j} v_i = \delta_{ij}. 
\end{equation}
The general solution can be given in terms of inverse Gram matrices~\cite{Lee:2013mka}, 
\begin{align}
    \frac{\partial}{\partial p_i \cdot p_j}=\sum_{k=1}^5 (\mathcal{G}(p_1,...,p_5)^{-1})_{kj}p_k\cdot \frac{\partial}{\partial p_i}
\end{align}
after the Mandelstam derivatives have been expressed by derivatives with respect to scalar products via the chain rule.  Then, acting with a momentum derivative merely raises some of the propagator powers, allowing us to use IBP identities to reduce the Mandelstam derivatives of our basis elements to linear combinations of master integrals according to
\begin{equation}
    \frac{\partial \tilde{I}_j}{\partial v_i} = \eps r_{ij}^k \tilde{I}_k.
\end{equation}
Importantly, the $r_{ij}^k$ are manifestly rational functions of the Mandelstam invariants. In principle, we could find the logarithmic PDE matrix $A$ by integrating the rational function $r_{ij}^k$ into logarithms. This would provide us with a constructive determination of the hexagon alphabet~$\mathbb{A}$. However, we find that it is more efficient to start with an educated ansatz for the alphabet and match it to the derivatives with respect to Mandelstam invariants entry by entry. 
In the following two sections, we describe how to construct the hexagon alphabet from the known alphabet for two--loop pentagon functions appended by a set of ten additional hexagon letters. 
 We record the expression for $A$ \eqref{eq:PDEMatrix} in terms of the alphabet letters to be described in the next section in the ancillary file \verb|PDEMatrix.txt|.
\subsection{The one--loop hexagon alphabet}
\subsubsection{Alphabet letters from one--mass five--point integrals}
\label{sec:AlphabetPent}
In this and the following section, we describe the hexagon alphabet $\mathbb{A}$ by listing the letters $W_i$ necessary to express the canonical differential equation matrix. The one--loop hexagon function alphabet has a large overlap with the two--loop one--mass five--point alphabet~\cite{Abreu:2020jxa, Chicherin:2021dyp}, since by contracting one of the propagators we retrieve the one--mass pentagon integral.

Hence, as a starting point, we can use the alphabet letters provided in the ancillary files of Ref.~\cite{Chicherin:2021dyp} and translate them to the six--point kinematic invariants. We obtain the relevant alphabet letters by treating the massive momenta as a sum of the two massless ones and using cyclic permutations to obtain the remaining combinations of external momenta. 

The relevant part of the function alphabet, retrieved in this way, has 93 alphabet letters needed to express the differential equation matrix $A$ up to and including all of the one--mass pentagon integrals. After describing the letters we obtain from this construction, in the next section we construct ten additional letters needed to express the full $A$ matrix for our set of master integrals.

We call a letter $W_i$ odd if $d \log W_i$ changes a sign when changing the sign of the square root. The square roots appearing in the function alphabet correspond to the inverse leading singularities defined in section \ref{subsec:leadingsingularities}.

The first 39 letters are purely rational and therefore even:
\begin{gather}
   W_1=v_1,\quad W_{i+1}=T^iW_1,\quad i=1,\ldots,5,\\
   W_7=v_7,\quad W_{i+7}=T^iW_7,\quad i=1,\ldots,2,\\
   W_{10}=v_1-v_7,\quad W_{i+10}=T^iW_{10},\quad i=1,\ldots,5,\\
   W_{16}=v_1-v_9,\quad W_{i+16}=T^iW_{16},\quad i=1,\ldots,5,\\
   W_{22}=-v_1-v_2+v_7,\quad W_{i+22}=T^iW_{22},\quad
   i=1,\ldots,5,\\
   W_{28}=v_1+v_4-v_7-v_9,\quad W_{i+28}=T^iW_{28},\quad i=1,\ldots,2,\\
   W_{31}=-v_1 v_4+v_7 v_9,\quad W_{i+31}=T^iW_{31},\quad i=1,\ldots,2,\\
   W_{34}=v_1(-v_3 +v_9) + v_9 (v_3-v_5-v_9),\quad W_{i+34}=T^iW_{34},\quad i=1,\ldots,5.
\end{gather}
We have another eight even letters containing just one of the square roots
\begin{gather}
    W_{41}=\sqrt{\Delta_3(v_1,v_3,v_5)},\quad W_{42}=TW_{41},\\
    W_{43}=\sqrt{\Delta_5(v_5,v_6,v_1,v_8,v_9,v_3)},\quad W_{i+43}=T^iW_{43},\quad i=1,\ldots,5.
\end{gather}
Next, we have 40 odd letters containing one square root and polynomials in kinematic invariants:
\begin{gather}
    W_{49}=\dfrac{-v_1+v_3 -v_5-\sqrt{\Delta_3(v_1,v_3,v_5)}}{-v_1+v_3 -v_5+\sqrt{\Delta_3(v_1,v_3,v_5)}},\quad W_{50}=TW_{49},\\
    W_{51}=\dfrac{v_1-v_3 -v_5-\sqrt{\Delta_3(v_1,v_3,v_5)}}{v_1-v_3 -v_5+\sqrt{\Delta_3(v_1,v_3,v_5)}},\quad W_{52}=TW_{51},\\
    W_{53}=\dfrac{v_1-v_3 +v_5 -2v_7-\sqrt{\Delta_3(v_1,v_3,v_5)}}{v_1-v_3 +v_5-2v_7+\sqrt{\Delta_3(v_1,v_3,v_5)}},\quad W_{i+53}=T^iW_{53},,\quad i=1,\ldots,5,\\
    W_{59}=\dfrac{v_2v_3-v_2v_5 -v_3v_7+v_7v_8+v_1(-v_2+v_8) -\sqrt{\Delta_5(v_1,v_2,v_3,v_7,v_8,v_5)}}{v_2v_3-v_2v_5 -v_3v_7+v_7v_8+v_1(-v_2+v_8) +\sqrt{\Delta_5(v_1,v_2,v_3,v_7,v_8,v_5)}},\\ W_{i+59}=T^iW_{59},,\quad i=1,\ldots,5,\\
    W_{65}=\dfrac{-v_2v_3-v_2v_5 +v_3v_7+v_7v_8+v_1(v_2-v_8) -\sqrt{\Delta_5(v_1,v_2,v_3,v_7,v_8,v_5)}}{-v_2v_3-v_2v_5 +v_3v_7+v_7v_8+v_1(v_2-v_8) +\sqrt{\Delta_5(v_1,v_2,v_3,v_7,v_8,v_5)}},\\ W_{i+65}=T^iW_{65},,\quad i=1,\ldots,5,\\
    W_{71}=\dfrac{v_2v_3+v_2v_5 -v_3v_7-2v_2v_8+v_7v_8-v_1(v_2+v_8) -\sqrt{\Delta_5(v_1,v_2,v_3,v_7,v_8,v_5)}}{v_2v_3+v_2v_5 -v_3v_7-2v_2v_8+v_7v_8-v_1(v_2+v_8) +\sqrt{\Delta_5(v_1,v_2,v_3,v_7,v_8,v_5)}},\\ W_{i+71}=T^iW_{71},,\quad i=1,\ldots,5,\\
    W_{77}=\dfrac{R_1 -\sqrt{\Delta_5(v_1,v_2,v_3,v_7,v_8,v_5)}}{R_1 +\sqrt{\Delta_5(v_1,v_2,v_3,v_7,v_8,v_5)}},\\
    W_{i+77}=T^iW_{77},,\quad i=1,\ldots,5,\\
    W_{83}=\dfrac{R_2 -\sqrt{\Delta_5(v_1,v_2,v_3,v_7,v_8,v_5)}}{R_2 +\sqrt{\Delta_5(v_1,v_2,v_3,v_7,v_8,v_5)}},\\ W_{i+83}=T^iW_{83},,\quad i=1,\ldots,5,
\end{gather}
where we used 
\begin{gather}
    R_1=v_2(-v_3+v_5-2v_7) + v_1 (v_2+2v_5-2v_7-v_8) + v_7(v_3-2v_5+2v_7+v_8),\\
    R_2=v_2(v_3+v_5-2v_8) + (2v_5-v_7-2v_8)(v_3-v_8)+v_1(-v_2+v_8).
\end{gather}

The last six letters involve two different square roots, therefore these are even under simultaneous change of both signs of square roots, but odd under the sign change of one of the square roots:
\begin{gather}
    W_{89}=\dfrac{R_3 -\sqrt{\Delta_3(v_1,v_3,v_5)\Delta_5(v_1,v_2,v_3,v_7,v_8,v_5)}}{R_3 +\sqrt{\Delta_3(v_1,v_3,v_5)\Delta_5(v_1,v_2,v_3,v_7,v_8,v_5)}},\\ W_{i+89}=T^iW_{89},,\quad i=1,\ldots,5,
\end{gather}
where
\begin{align}
    R_3=&v_1^2(v_2-v_8)+v_1(-2v_2(v_3+v_5)+v_3(-2v_5+v_7)+(v_3+v_5+v_7)v_8)\notag\\
    &+(v_3-v_5)(v_2(v_3-v_5)+v_7(-v_3+v_8)).
\end{align}

\subsubsection{One--loop hexagon letters}
\label{sec:AlphabetHex}
The additional ten letters, appearing for the first time in the hexagon integral, are obtained following the construction of one--loop alphabet letters from \cite{Chen:2022fyw}. The letters involving square roots can be written in the form
\begin{equation}
    \dfrac{P(\Vec{v})-\sqrt{Q(\Vec{v})}}{P(\Vec{v})+\sqrt{Q(\Vec{v})}}
\end{equation}
where $P$ and $Q$ are polynomials in the kinematic variables $v_i$. These polynomials are expressed in terms of Gram determinants defined in (\ref{eq:GramDet}).

We construct one even letter
\begin{equation}
    W_{40}
    = \dfrac{- \Delta_6}{G(p_1,p_2,p_3,p_4,p_5)},
\end{equation}
three odd letters
\begin{equation}  \label{eq:hextobox}
    W_{101}
    =\dfrac{v_3v_6v_7-v_1v_4v_8-v_2v_5v_9+v_7v_8v_9-\sqrt{\Delta_6}}{v_3v_6v_7-v_1v_4v_8-v_2v_5v_9+v_7v_8v_9+ \sqrt{\Delta_6}},
\end{equation}
\begin{equation}
    W_{i+101}=T^i W_{101},\quad i=1,\ldots,2,
\end{equation}
and six letters involving two square roots
\begin{equation}
    W_{95}
    =\dfrac{R_4-\sqrt{\Delta_5(v_1,v_2,v_3,v_7,v_8,v_5)\Delta_6}}{R_4+\sqrt{\Delta_5(v_1,v_2,v_3,v_7,v_8,v_5)\Delta_6}},
\end{equation}
\begin{equation}
    W_{i+95}=T^i W_{95},\quad i=1,\ldots,5,
\end{equation}
where 
\begin{align}
   R_4=&v_1^2v_4v_8(-v_2+v_8) - (v_2(v_3-v_5)+v_7(-v_3+v_8))(v_3v_6v_7+(v_2v_5-v_7v_8)v_9) \notag\\
   &+v_1(v_2^2v_5(2v_3-v_9)+v_7v_8(v_3(v_4+v_6)-v_8(v_4+v_9)) \notag\\
   &+v_2(v_3(-2v_5v_6+v_7(v_6-2v_8)+v_4(-2v_5+v_8))+v_8(v_4v_5+(v_5+v_7)v_9))).
   \label{eq:R4Expr}
\end{align}

Note that we can relate the hexagon square root to the one known from the study of dual conformal hexagon integrals \cite{Dixon:2011ng}. Alphabet letters in this case are expressed in terms of three dual conformally invariant cross--ratios
\begin{equation}
    u_1=\dfrac{v_1v_4}{v_7v_9}, \quad u_2=\dfrac{v_2v_5}{v_7v_8}, \quad u_3=\dfrac{v_3 v_6}{v_8 v_9},
\end{equation}
and 
\begin{equation}
    \Delta=(1-u_1-u_2-u_3)^2-4u_1u_2u_3.
\end{equation}
It is easy to verify that the relation between the hexagon square root $\sqrt{\Delta_6}$ and $\sqrt{\Delta}$ is
\begin{equation}
    \sqrt{\Delta_6}=v_7v_8v_9 \sqrt{\Delta}.
\end{equation}

The alphabet letters are related to the possible singularities of the Feynman integrals. Therefore, they can be deduced from the solutions of the Landau equations~\cite{Landau:1959fi}. In practice finding these solutions is not an easy task but in the setting of $n$--gon one--loop Feynman integrals they can be described in closed form. These solutions can be computed using the Cayley matrix and the associated determinant, see Ref.~\cite{Hannesdottir:2022bmo,Griffon:1969}
\begin{equation}
    C= \begin{pmatrix}
    0 & \vec{1}^T \\
    \vec{1} & x_{ij}^2
    \end{pmatrix}, \quad i,j=0,1,\dots,n
\end{equation}
where $\vec{1}$ denotes a $(n+1)$--vector of 1's and $x_{ij}=x_i - x_j$ are given in terms of dual coordinates $x_i$. Here the coordinate $x_{0}$ is related to the loop momentum $l$ and the external momenta are related to dual coordinates as $p_i=x_i-x_{j+1}$ with $x_{i+n}=x_i$. For the one--loop hexagon, the Cayley matrix is an $8 \times 8$ matrix whose determinant equals $\Delta_6$. While we indeed observe that the different minors of the Cayley matrix correspond to different polynomials which appear in the alphabet letters, we are not aware of an explicit algorithm that constructs the alphabet from the Cayley matrix only.

To summarise, the one--loop hexagon alphabet $\mathbb{A}$ contains 103 letters that are algebraic functions of the kinematic invariants $v_i$. The square roots $\sqrt{\Delta_3}$, $\sqrt{\Delta_5}$ and $\sqrt{\Delta_6}$, appearing in the alphabet letters $W_i$, correspond to the inverse leading singularities of three--mass triangles, one--mass pentagons and hexagon, respectively. We classify the alphabet letters according to parity, therefore we have 48 even letters and 55 odd letters, 12 of which contain two square roots.
We stored the alphabet in the file \verb|HexagonAlphabet.txt| and the  associated square roots in \verb|HexagonSqrts.txt|.

\section{Analytic result from differential equations}
\label{sec:DE}
In this section we describe the solution to the canonical differential equation and introduce the notion of the symbol. At the symbol level, we verify the extended Steinmann relations. To fully determine the solution to the canonical differential equation, we fix the boundary constants by imposing the absence of spurious singularities and by matching to the analytic solution of a bubble integral at the boundary point. With the knowledge of the boundary values up to order $\mathcal{O}(\epsilon^4$), we obtain one--fold integral representation for numerical evaluation which we validate at several points. We study the symbol of the one--loop hexagon integral at order $\mathcal{O}(\epsilon)$. Finally, we study the limit of four--dimensional external states and dependencies between the letters at the level of differential equations. 

\subsection{Solving the canonical differential equations}

Formally, the solution to the canonical differential equation (\ref{eq:CanonicalPDE}) can be written as a path ordered integral 
\begin{equation}
    \tilde I (\vec{v},\eps) = \mathbb{P} \exp (\eps \int_{\gamma} \text{d}A) \cdot \vec{b}, 
\end{equation}
where $\gamma$ denotes a path from a boundary point $\vec v_0$ to $\vec v$ and $\vec b = \tilde I (\vec v_0)$ is a vector of master integrals values at the boundary point $\vec v_0$. This equation is to be understood as a Laurent expansion around $\epsilon=0$,
\begin{equation}
    \tilde I (\vec{v},\eps) = \sum_{k=0}^{\infty} \eps^k \tilde{I}^{(k)}(\vec v).
\end{equation} 
Substituting this expansion in the canonical differential equation (\ref{eq:CanonicalPDE}), we see that the equation decouples order by order in $\epsilon$
\begin{equation}
    d \tilde{I}^{(k+1)}(\vec v)= \sum_{k \ge 0} (d A (\vec v)) \tilde{I}^{(k)}(\vec v)
\end{equation}
The $(k+1)$--th term in the expansion is then given by a $(k+1)$--fold iterated integral along the contour $\gamma$ of the matrix differential form $A$
\begin{equation}
    \tilde{I}^{(k+1)}(\vec v) = \int_{\gamma} (d A (\vec v)) \tilde{I}^{(k)}(\vec v) + \vec{b}^{(k)},
\end{equation}
where $\vec{b}^{(k)}$ are weight $k$ boundary values which we discuss in the following section. We can rewrite this equation in terms of Chen's iterated integrals \cite{Chen:1977oja}
\begin{equation}
    \tilde{I}^{(k)} (\vec{v}) = \sum_{k'=0}^{k} \sum_{i_1,\dots, i_{k'} \in \mathbb{A}} a^{(i_1)} \dots a^{(i_{k'})} \vec{b}^{(k)} \left[W_{i_1},\dots,W_{i_{k'}} \right]_{\vec{v}_0} (\vec{v}),
    \label{eq:solutionChen}
\end{equation}
where 
\begin{equation}
    \left[W_{i_1},\dots,W_{i_{k}} \right]_{\vec{v}_0} (\vec{v})=\int_{\gamma} \mathrm{d} \log W_k (\vec{v}')  \left[W_{i_1},\dots,W_{i_{k-1}} \right]_{\vec{v}_0},
    \label{eq:ChenIter}
\end{equation}
and $\left[\; \right]_{\vec{v}_0}=1$. In order for the integral to be well--defined, it has to be homotopy invariant i.e.~it should be independent of the choice of the contour as long as the singularities are not crossed. Since the differential equation satisfies the integrability conditions\footnote{Integrability conditions for the differential equation require that $\partial_{v_j}A_i - \partial_{v_i}A_j=0$ and $\left[A_i,A_j\right]=0$.}, the solution (\ref{eq:solutionChen}) is homotopy invariant, but the separate integrals (\ref{eq:ChenIter}) are not.

Furthermore, we can introduce the notion of the symbol which is a useful tool for studying the polylogarithmic functions (see ref. \cite{Goncharov:2010jf}). It maps a $k$--fold iterated integral to the $k$--fold tensor product
\begin{equation}
    \mathcal{S} \left(\left[W_1,\dots,W_{k} \right]_{\vec{v}_0} (\vec{v}) \right) = W_1 \otimes \dots \otimes W_k.
\end{equation}
Additionally, we will use $\mathcal{S}^{(k)}(f)$ to denote the symbol of the weight--$k$ part of $f$, i.e.~if $f$ has an expansion
\begin{equation}
    f = \sum_k \epsilon^k f_k
\end{equation}
then we define
\begin{equation}
    \mathcal{S}^{(k)}(f) = \mathcal{S}(f_k).
\end{equation}

\subsection{Steinmann relations}
One immediate consequence we can draw from the form of $A$ is the validity of the extended Steinmann relations  \cite{Caron-Huot:2016owq,Caron-Huot:2019bsq} for all integrals in the one-loop hexagon family. For this family, these relations require that the three--point Mandelstam variables $s_{123}$, $s_{234}$ and $s_{345}$ never appear as consecutive letters in the symbol. Physically, they reflect the incompatibility of the different three--particle cuts on all possible Riemann--sheets. The differential equation matrix $A$ guarantees that the relations hold at any order in $\epsilon$ at any depth into the symbol by the simple identities \cite{Chicherin:2020umh}
\begin{equation}
    c_{jk}^a c_{kl}^b = 0,
    \label{eq:SteinmannIdentity}
\end{equation}
for all $j,l \in \{1,...,33\}$ and for $a\neq b$, $a,b \in \{7,8,9\}$, which is straightforward to verify. Of course, these are not the only adjacency relations that follow from the differential equation. In fact, all letters $W_a, W_b$ for which an identity of the form \eqref{eq:SteinmannIdentity} is satisfied will never appear next to each other in the symbol at any order in $\epsilon$. It will be very interesting to investigate which of these relations continue to hold at the two--loop level and at higher loops.
\subsection{Absence of spurious singularities fixes the boundary constants}
\label{sec:BoundaryConstants}
\begin{figure}
    \centering
    \includegraphics[width=\textwidth]{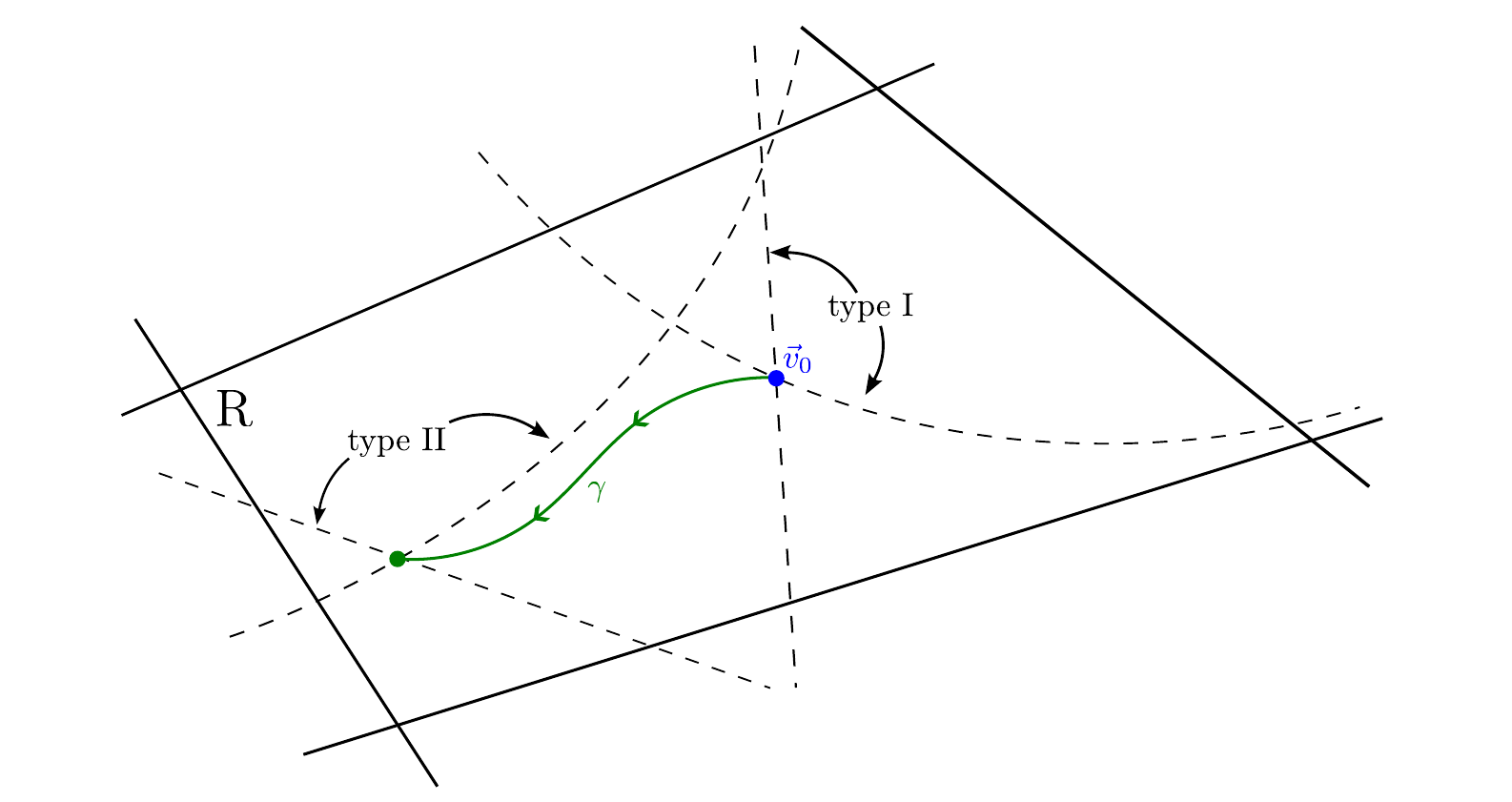}
    \caption{Graphical representation of the kinematic region R which is bounded by the solid black lines. The dashed lines represent singular hypersurfaces on which some of the letters diverge. To fix the boundary constants, we start at some point $\vec{v}_0$ and impose cancellation of all singularities in the PDE matrix at this point (which we call type I singularities). Then, we set up one--parameter paths labelled by $\gamma$ that lead to all other singular hypersurfaces. Imposing cancellation of divergencies at the end--points of the path (which we call type II singularities) further restricts the boundary constants.}
    \label{fig:SingularityCartoon}
\end{figure}
To fully determine the solution to \eqref{eq:CanonicalPDE} we need to supplement it with boundary information, i.e.~we need to provide the analytical values for our basis of master integrals at a single point in the space of kinematical variables. We will see in this section, that we can determine the values of our integral basis at the point
\begin{equation}
     \vec v_0 = \{-1,-1,-1,-1,-1,-1,-1,-1,-1\}
\end{equation}
up to the order $\epsilon^4$ by imposing the absence of certain spurious singularities and by matching to the analytical solution of a bubble integral. To be a bit more explicit, even though the Feynman integrals in our basis are manifestly finite all throughout the Euclidean region, this is not true for the PDE matrix $A$. In fact, the vanishing (or divergence) of any the hexagon letters
\begin{equation}
    W_i = 0, 
\end{equation}
defines a hypersurface that might intersect the Euclidean region. The situation is sketched in Fig.~\ref{fig:SingularityCartoon}. Imposing finiteness of the solution to \eqref{eq:CanonicalPDE} on any of these hypersurfaces severly constrains the possible boundary values at the point $\vec v_0$, c.f.~\cite{Chicherin:2018mue}.
Hence, we will determine the $c_{ij}$ in the expansion
\begin{equation}
    \tilde I_i(\vec v_0) = \sum_{j=0}^4 c_{ij} \epsilon^j.
\end{equation}
We also denote the vector of boundary constants as 
\begin{align}
    \vec b = \tilde I(\vec v_0).
\end{align}
In the following analysis, we will call spurious letter singularities at the bulk point $\vec{v}_0$ type--I singularities and singularities on any other hypersurfaces that do not contain $\vec{v}_0$  type--II singularities. 
\subsubsection{Bubble integral}
The bubble integrals can be calculated analytically, giving e.g.\ 
\begin{align}
   \tilde{I}_1 &= \epsilon e^{\epsilon \gamma_E}\frac{\Gamma(-\epsilon)^2 \Gamma(1 + \epsilon)}{\Gamma(-2 \epsilon)(-v_1)^\epsilon}
\end{align}
Hence, at the boundary point, we get
\begin{align}
    \tilde{I}_1 (\vec v_0)= -2 + \zeta_2 \epsilon^2 + \frac{14}{3} \zeta_3 \epsilon^3 + \frac{47}{20} \zeta_2^2 \epsilon^4+  \mathcal{O}(\epsilon^5).
\end{align}
\subsubsection{Absence of Type--I singularities}
To remove the spurious letter divergences at the point $\vec{v}_0$, we parametrise paths that start at this point by an infinitesimal parameter $\delta$. Then, the divergent letters will have an expansion in $\delta$ according to 
\begin{align}
    \log W_j = c_j \log \delta + \mathcal{O}(1). 
\end{align}
The set of letters vanishing at $\vec{v}_0$ is given by
\begin{equation}
    \mathbb{A}_0 = \{W_{10},\dots,W_{21},W_{28},\dots,W_{33}\}.
\end{equation}
Since the master integrals are supposed to diverge only on the boundaries of the Euclidean region, the boundary vector must be such that 
\begin{align} 
 \lim_{\delta \to 0} A \vec b = \text{finite}.
\end{align}

Using different parametrizations to approach the boundary point on different curves, we find that this requirement constrains $\vec b$ (non-perturbatively in $\epsilon$) to be of the form
\begin{align}
    \vec b = \{&b_1,b_1,b_1,b_1,b_1,b_1,b_1,b_1,b_1,b_{10},b_{11},b_{12},b_{13},b_{14},b_{15},b_{16},b_{17},b_{18},b_{19},b_{20},b_{21},\notag\\&-b_{12}+b_{14}-b_{15}+b_{17}+b_{18}-b_{19}+b_{21}, -b_{13}+b_{14}-b_{16}+b_{17}+b_{18}-b_{20}+b_{21},\notag\\&0,0,0,0,0,0,0,0,0,0\}.
\end{align}
\subsubsection{Absence of Type--II singularities}
Beyond the singular letters at the point $\vec v_0$, there are additional letters that diverge at some other point in the Euclidean region, e.g.\ 
\begin{equation}
    \log W_{21} = \log(-v_1-v_2+v_7).
\end{equation}
To ensure that these singularities do not actually show up in our solution of the differential equation, we integrate it from the boundary point $\vec v_0$ to the hypersurface where these additional letters diverge. Then, imposing the absence of logarithmic divergences further constrains the boundary constants $b_i$. Since we generally do not have access to a non--perturbative solution of \eqref{eq:CanonicalPDE}, these constraints are perturbative in $\epsilon$. 
A convenient path from $\vec v_0$ to the hypersurface where $W_{21}$ vanishes is given by
\begin{align}
    \vec v = \{-1,-(1-x)^2, -1, -(1-x)^2, -1, -(1-x)^2, -1+x, -1+x, -1+x\},
    \label{eq:path}
\end{align}
which satisfies the Gram constraint. 
On this line, the hexagon alphabet simplifies to 
\begin{align}
    \mathbb{A}\rightarrow  \mathbb{A}_\text{line} = \{x, 1-x, x - \rho, x - \bar \rho\},
    \label{eq:reducedAlpha}
\end{align}
with $\rho = \frac{1}{2}(1 + i \sqrt{3})$.
Hence there are four potential singular points on the line. The absence of type--I singularities ensures that all the $\log x$ singularities drop out. The $\log(1-x)$ singularities lie on the boundary of the Euclidean region and hence they are not necessarily unphysical. Finally, the singularities at the points $\rho$ and $\bar \rho$ are non--physical and we can further determine the boundary values by imposing that they drop out.

Using \PLT~\cite{Duhr:2019tlz}, it is straightforward to integrate the differential equation with the alphabet~\eqref{eq:reducedAlpha} to Goncharov--Polylogarithms~\cite{Goncharov:2011abc} with entries $\{0, 1, \rho, \bar \rho\}$. To isolate the singularities, we use shuffle identities to bring the $G$--functions to a convenient Lyndon basis~\cite{RADFORD1979432}. Then, imposing that all $\log(x-\rho)$ drop out for $x\rightarrow \rho$ (and similarly for $x\rightarrow \bar \rho$), we can fully determine the boundary values order by order. Up to $\mathcal{O}(\epsilon^4)$, we find the following non--vanishing boundary constants,\footnote{These constants are  embedded in the basis of polylogarithms at sixth roots of unity discussed in~Ref.~\cite{Henn:2015sem}.}
\begin{align}
b_{1,...,9} &= -2 + \zeta_2  \epsilon^2 + \frac{14}{3} \zeta_3 \epsilon^3 + \frac{47}{8} \zeta_4 \epsilon^4 + \mathcal{O}(\epsilon^5),\notag\\
b_{10,11}&= i(6 \text{ Im}[\text{Li}_2(1-\rho^2)+2\pi \log{3})\epsilon^2\notag\\
&\quad + i\Big(\frac{10}{9}\pi \zeta_2+12  \text{ Im}[\text{Li}_3(1-\rho^2)]+\pi  \log (3)^2\Big)\epsilon^3 \notag\\
&\quad + i\Big(-3 \zeta_2 \text{ Im}[\text{Li}_2(1-\rho^2)]+24 \text{ Im}[\text{Li}_4(1-\rho^2)]\notag\\
&\qquad\quad +\frac{16}{3} \text{ Im}[\text{Li}_4(1+\rho^2)]+\frac{\pi}{3} \log(3)^3 + 12 \pi \zeta_2 \log(3)\Big)\epsilon^4 + \mathcal{O}(\epsilon^5),\notag\\
    b_{12,...,17} &= 2 - 3\epsilon^2  \zeta_2 - \frac{20}{3} \epsilon^3 \zeta_3 - \frac{43}{8}\epsilon^4  \zeta_4 + \mathcal{O}(\epsilon^5),\notag\\
     b_{18,...,23} &= 1 - \epsilon^2 \frac{\zeta_2}{2} + \epsilon^3 \frac{\zeta_3}{3}  +\epsilon^4 \Big(\frac{17}{16}\zeta_4+6 \text{ Im}[\text{Li}_2(1-\rho^2)]^2 \notag\\
    &\qquad \ + \frac{2}{3}\pi^2 \log 3^2 + 4\pi\log 3 \text{ Im}[\text{Li}_2(1-\rho^2)]\Big) + \mathcal{O}(\epsilon^5).
    \label{eq:BoundaryValues}
\end{align}
In particular, the inverse leading singularities of all pentagon integrals and the hexagon integral vanish on the entire line, leading to trivial boundary conditions for these functions. 
With these constants determined, we actually have full analytic control over the entire function space at weight four on the line given by \eqref{eq:path}. We have used this to numerically confirm our calculation via \verb|AMFlow|~\cite{Liu:2022chg}, see Table \ref{tab:numerics}.

The Boundary constants in eq.~\eqref{eq:BoundaryValues} are valid in the Euclidean region. To obtain the boundary constants in the physical region, we can either perform an analytic continuation or determine the boundary constants directly in the physical region of interest following a similar procedure~\cite{Chicherin:2018mue}.

\subsection{Integral representations for numerical evaluation}\label{subsec:intrepw4}

Now that we fixed the boundary values up to order $\mathcal{O}(\eps^4)$, we can write down the solution to the differential equation (\ref{eq:CanonicalPDE}) order by order in $\eps$. 

If we know our basis integrals at a single point $v_i(0)$ up to weight four,
\begin{equation}
    \tilde{I}(v_i(0)) = \tilde{I}_0 = \sum_{k=0}^4 \epsilon^k \tilde{I}_0^{(k)} + \mathcal{O}(\epsilon^5)
\end{equation}
and want to evaluate them numerically at some other point $v_i(1)$, we can use a trick to rewrite the solution to the differential equation in terms of a one--fold integration over the weight--two functions \cite{Gehrmann:2018yef,Caron-Huot:2014lda} from the section \ref{subsec:fweighttwo}, which is suitable to numerical integration. 
We set up a straight line between our starting and endpoint
\begin{equation}
    v_i(t) = (1-t)v_i(0) + t v_i(1).
\end{equation}
Then, we can explicitly write the solution to the canonical differential equation as
\begin{align}
    \tilde{I}(t) = \mathcal{T}\exp{[\epsilon \int_0^t \text{d}t' \frac{\text{d}A}{\text{d}t'}]}\tilde{I}_0,
\end{align}
with the time--ordering operator $\mathcal{T}$. 
Hence, the weight--$k$ solution at the end--point $v_i(1)$ reads
\begin{equation}
    \tilde{I}^{(k)}(1)=\tilde{I}^{(k)}_0 + \sum_{i=1}^k \prod_{j=1}^i \int_0^{t_{j-1}} \text{d} t_j \frac{\text{d}A}{\text{d}t_j} \tilde{I}_0^{(k-i)},
    \label{eq:SolUpToWeight4}
\end{equation}
where $t_0=1$.
By matching towards the expansions in \eqref{eq:SolUpToWeight4}, we find
\begin{align}
    \tilde{I}^{(1)}(t) &= \tilde{I}_0^{(1)} + \int_0^t \text{d}t_1 \frac{\text{d}A}{\text{d}t} \tilde{I}^{(0)},\notag\\
    \tilde{I}^{(2)} (t) &= \tilde{I}_0^{(2)} + \int_0^t \text{d}t_1 \frac{\text{d}A}{\text{d}t} \tilde{I}^{(1)}_0 + \int_0^t \text{d}t_1 \int_0^{t_1} \text{d}t_2 \frac{\text{d}A}{\text{d}t_1}\frac{\text{d}A}{\text{d}t_2} \tilde{I}^{(0)}_0,
\end{align}
where $\tilde{I}^{(1)}(t)$ and $\tilde{I}^{(2)}(t)$ denote the weight--one and weight--two solution to the canonical differential equation (\ref{eq:CanonicalPDE}) at the point $t$. These solutions can be rewritten in terms of the basis of functions introduced in section \ref{subsec:fweighttwo}. At weight one, we know that the only function that can appear is the logarithm
\begin{equation}
    f_i^{(1)}=\left[W_i \right]=\log(-v_i), \quad i=1,\dots, 9,
\end{equation}
while at weight two we can have products of logarithms alongside dilogarithms. For example, a weight--two function appearing in the solution for box integrals is
\begin{equation}
    f_1^{(2)}= - \left[ \frac{W_1}{W_7}, \frac{W_{10}}{W_7} \right] = \mathrm{Li}_2 \left(1-\dfrac{v_1}{v_7}\right),
\end{equation}
and remaining functions can be expressed in similar manner in terms of iterated integrals. 
Therefore the solution up to weight two for any integral in our basis is a linear combination of these functions. 

At weight three, the solution is expressed as a one--fold integral over the weight--two functions $\vec{f}^{(2)}$
\begin{equation}
    \tilde{I}^{(3)}(1)= \tilde{I}_0^{(3)} +   \int_0^1 \text{d}t_1 \frac{\text{d}A}{\text{d}t_1} \vec{f}^{(2)}(t_1)
\end{equation}
whereas at weight four we have
\begin{align}
    \tilde{I}^{(4)}(1) &= \tilde{I}_0^{(4)} + \int_0^1 \text{d}t \frac{\text{d}A}{\text{d}t} \tilde{I}_0^{(3)} +\int_0^1 \text{d}t_1 \int_0^{t_1}\text{d}t_2 \frac{\text{d}A}{\text{d}t_1}\frac{\text{d}A}{\text{d}t_2}\vec f^{(2)}(t_2)\notag\\
    &= \tilde{I}_0^{(4)} + \int_0^1 \text{d}t \left(\frac{\text{d}A}{\text{d}t} \tilde{I}_0^{(3)} + \left(A(1)-A(t)\right)\frac{\text{d}A}{\text{d}t} \vec f^{(2)}(t)\right),
\end{align}
where we used integration by parts with respect to $t_1$ to reduce the last term in the first line to a one--fold integral.

Once we have analytical results up to weight two and integral representation up to weight four in the dimensional regulator we can validate our computation numerically. We choose several points in the bulk of the Euclidean region (see Table \ref{tab:numerics}). The point $\vec{v}^{(1)}$  corresponds to the point $K^{(3)}$ from Ref.~\cite{Bern:2008ap} while the points $\vec{v}^{(2)}$ and $\vec{v}^{(3)}$ do not satisfy the Gram determinant constraint and therefore do not satisfy the four--dimensional kinematics. Numerical values obtained using the one--fold integral representation are validated via \verb|AMFlow| up to the desired precision of 20 relevant digits where we find a complete agreement. 
\begin{table}[t!]
    \centering
    \begin{tabular}{ | c | c |  c | } 
    \hline
    kinematic point     & $\Bar{G}_5=0$  &  $\tilde{I}_{33}$ \\
    \hline
    \rule{0pt}{6ex} $\Vec{v}^{(1)}=\left \lbrace-1,-1,-1,-1,-1,-1,-\frac{1}{2},-\frac{5}{8},-\frac{17}{14}\right\rbrace$ & $\checkmark$ & \parbox[c]{5cm}{$1.69878610466574714627 i \eps^3 \\+ 6.62873216549319714468 i \eps^4 \\+ \mathcal{O}(\eps^5)$} \\[0.55cm]
    \hline
    \rule{0pt}{6ex} $\Vec{v}^{(2)}=\left \lbrace-\frac{2}{3},-\frac{7}{10},-\frac{9}{11},-\frac{15}{17},-\frac{24}{29},-\frac{30}{37},-\frac{37}{43},-\frac{47}{53},-\frac{53}{59}\right\rbrace$& $\times$& \parbox[c]{5cm}{$ 1.2966474952363382027 i \eps^3 \\+ 5.241756401399539064 i \eps^4 \\+ \mathcal{O}(\eps^2)$}  \\[0.55cm]
    \hline\rule{0pt}{6ex} $\Vec{v}^{(3)}=\left \lbrace-\frac{7}{9},-\frac{4}{5},-\frac{29}{33},-\frac{47}{51},-\frac{77}{87},-\frac{97}{111},-\frac{39}{43},-\frac{49}{53},-\frac{55}{59}\right\rbrace$ & $\times$ & \parbox[c]{5cm}{$0.81389548925976547185 i \eps^3 \\+ 3.2221858302838235961 i \eps^4 \\+ \mathcal{O}(\eps^5)$} \\[0.55cm]
    \hline
    \end{tabular}
    \caption{Numerical evaluation of the hexagon integral at several kinematic points. The values correspond to the UT hexagon integral and are obtained using the one--fold integral representation.}
    \label{tab:numerics}
\end{table}

Note that a subset of letters diverges in the $D_{ext} \to 4$ limit, as we discuss in the section~\ref{subsec:limit4d}. These letters appear in the last row of the PDE matrix $A$ and therefore influence the numerical integration of the hexagon integral. We can deform the Mandelstam invariants by a small number 
to avoid numerical instabilities at the points satisfying the Gram determinant constraint. Then we can use the one--fold integral representation to numerically evaluate the hexagon integral. 

\subsection{The one--loop hexagon integral at order $\mathcal{O}(\eps)$}
Of course, the most interesting integral in our basis is the massless hexagon integral. In this section, we discuss its finite part and its order epsilon part in generic external spacetime dimensions $D_\text{ext}$. The finite part of the hexagon integral was first calculated in Ref.~\cite{Dixon:2011ng}. Using the canonical differential equation matrix $A$ and the boundary constants determined in section~\ref{sec:BoundaryConstants}, it is straightforward to determine the symbols for the $\mathcal{O}(1)$ and $\mathcal{O}(\epsilon)$ parts of the hexagon integral. In agreement with Ref.~\cite{Dixon:2011ng}, we find
\begin{equation}
  \mathcal{S}^{(3)}(\tilde{I}_{33}) = (u_1 \otimes u_2+u_2 \otimes u_1 - \sum_{j=1}^3 u_j \otimes (1-u_j))\otimes y_3 + (\text{cyclic})
\end{equation}
where we used the dual--conformal cross ratios
\begin{equation}
    u_1 = \frac{v_1 v_4}{v_7v_9}, \quad u_2 = \frac{v_2 v_5}{v_8 v_7}, \quad u_3 = \frac{v_3 v_6}{v_9 v_8},
\end{equation}
as well as the parity--odd dual--conformal letters
\begin{equation}
    y_1 = \frac{1+u_1-u_2-u_3-\sqrt{\Delta}}{1+u_1-u_2-u_3+\sqrt{\Delta}}, \quad y_2 = \frac{1+u_2-u_3-u_1-\sqrt{\Delta}}{1+u_2-u_3-u_1+\sqrt{\Delta}}, \quad y_3 = \frac{1+u_3-u_1-u_2-\sqrt{\Delta}}{1+u_3-u_1-u_2+\sqrt{\Delta}}
\end{equation}
both of which form threefold orbits under cyclic permutations of the external points. 

At weight four, the symbol of the hexagon decomposes according to\footnote{We note that $W_{101} = 1/y_3$, $W_{102} = 1/y_1$ and $W_{103} = 1/y_2$.}
\begin{align}
    \mathcal{S}^{(4)}(\tilde{I}_{33}) =&\  \mathcal{S}^{(3)}(\tilde{I}_{33})\otimes W_{40} + \sum_{i=1}^6 T_i\left( \mathcal{S}^{(3)}(\tilde{I}_{27})\otimes W_{96}\right) + \frac{1}{2}\sum_{i=1}^3 T_i \left(\Omega \otimes W_{101}\right),
\end{align}
where we remind the reader that $\tilde{I}_{27}$ and its permutations are the one--mass pentagons and $\Omega$ is given by the following combination of box integrals 
\begin{align}
    \Omega =&\  \mathcal{S}^{(3)}(\tilde{I}_{18})+\mathcal{S}^{(3)}(\tilde{I}_{19})+\mathcal{S}^{(3)}(\tilde{I}_{21})+\mathcal{S}^{(3)}(\tilde{I}_{22})\notag\\ &\ -\mathcal{S}^{(3)}(\tilde{I}_{13})-\mathcal{S}^{(3)}(\tilde{I}_{16})-\mathcal{S}^{(3)}(\tilde{I}_{24})-\mathcal{S}^{(3)}(\tilde{I}_{25})-\mathcal{S}^{(3)}(\tilde{I}_{26}).
    \label{eq:BoxCombination}
\end{align}
By inspecting these two expressions, we can determine the reduced alphabets that are necessary to describe the hexagon integral at weight three and weight four respectively. They read
\begin{equation}
    \mathbb{A}_{\text{hex}}^{(3)} = \{u_1,u_2,u_3,1-u_1,1-u_2,1-u_3, y_1,y_2,y_3\}
\end{equation}
and 
\begin{align}
\mathbb{A}_{\text{hex}}^{(4)} = \mathbb{A}\backslash \{W_{41},\dots, W_{48}\}.
\end{align}
Starting at weight five, all 103 letters of the one--loop hexagon alphabet show up in the symbol.
\subsection{Limit of four--dimensional external states}
\label{subsec:limit4d}

It is well known that for four--dimensional external momenta, the hexagon integral can be decomposed into a linear combination of pentagon integrals \cite{Melrose:1965kb,Binoth:1999sp},~i.e.~
\begin{equation}
    I_{33} = -\frac{1}{2}\sum_{j,k=1}^6 (S^{-1})_{jk} I_{26+j},
    \label{eq:HexIdentity4D}
\end{equation}
with
\begin{equation}
    S = -\frac{1}{2} \begin{pmatrix}
     0 & 0 & s_{45} & s_{123} & s_{23} & 0\\
     0 & 0 & 0 & s_{56} & s_{234} & s_{34}\\
     s_{45} & 0 & 0 & 0 & s_{61} & s_{345}\\
     s_{123} & s_{56} & 0 & 0 & 0 & s_{12}\\
     s_{23} & s_{234} & s_{61} & 0 & 0 & 0\\
     0 & s_{34} & s_{345} & s_{12} & 0 & 0\\
    \end{pmatrix}.
\end{equation}
In this section we show that this identity comes out of the differential equation for free as a finiteness condition in four--dimensional kinematics. 
When the kinematics are generated by four--dimensional vectors, the nine Mandelstam invariants are not independent anymore but they satisfy the Gram determinant constraint, 
\begin{equation}
    G(p_1,p_2,p_3,p_4,p_5) = 0.
\end{equation}
To solve the constraint we use the explicit momentum twistor parametrisation in Eq.~\eqref{eq:MomTwistRepl}. In terms of the momentum twistor variables, the leading singularities of the pentagons and the hexagons become perfect squares. However, substituting the parametrisation into the PDE matrix $A$, a subset of the logarithmic letters diverges, namely
\begin{equation}
    \mathbb{A}_\text{div} = \{W_{40}, W_{95},\dots, W_{100}\}.
\end{equation}
It is obvious why $W_{40}$ diverges, since its denominator is given by the Gram determinant. However, the divergence of the logs of the letters $W_{95}, \dots, W_{100}$ are less straightforward and require some explanation. Let us focus on $W_{95}$ since the behaviour of the other letters follows from cyclic permutations. It takes the form
\begin{align}
    W_{95} = \frac{R_4 - \sqrt{\Delta_{5,1}\Delta_6}}{R_4+\sqrt{\Delta_{5,1}\Delta_6}},
\end{align}
where 
\begin{equation}
    \Delta_{5,1} = \Delta_5(v_1,v_2,v_3,v_7,v_8,v_5)
\end{equation}
and $R_4$ is given in \eqref{eq:R4Expr}. By inspection, one finds
\begin{equation}
    \Delta_{5,1}\Delta_6 = R_4^2 - 4 v_1 v_2 v_3 (v_2 v_5 - v_7 v_8) G(p_1,p_2,p_3,p_4,p_5)
\end{equation}
which is valid for arbitrary $D$--dimensional kinematics. Clearly, in the limit of four--dimensional external kinematics, the product of $\Delta_{5,1}$ and $\Delta_6$ approaches the absolute square of $R_4$ and we have
\begin{equation}
    W_{95}\Big|_{D_\text{ext}\rightarrow 4}\rightarrow \frac{R_4 - |R_4|}{R_4 + |R_4|}.
\end{equation}
Depending on the sign of $R_4$, the letter $W_{95}$ will either vanish or diverge in the limit $D\rightarrow 4$, hence its logarithm will go to $\mp\infty$.

The divergent letters appear only in the last row of the matrix, i.e.~the derivative of the hexagon integral and they correspond to the coefficients of the pentagon and hexagon integrals in this derivative. Hence, for the four--dimensional limit of the hexagon integral to be finite, these divergences must cancel. To approach the surface where the Gram determinant constraint holds more carefully, we deform the two--particle Mandelstam invariants by a small parameter $\delta$, i.e.~
\begin{equation}
    v_i \rightarrow v_i + \delta, \quad i=1,\dots,6,
\end{equation}
explicitly breaking the Gram determinant constraint at order $\mathcal{O}(\delta)$. 
Then, the divergent letters take the form\footnote{Note that this analysis is valid for a generic point in four--dimensional kinematics, at which no other letters diverge. At more special points, there can be additional constraints emerging from the differential equation.}
\begin{equation}
    W_i = -c_i\log(\delta) + \mathcal{O}(1), \quad W_i \in \mathbb{A}_\text{div},
\end{equation}
where the signs of the $\log(\delta)$ divergence depend of the sign of $R_4$, i.e.
\begin{equation}
    c_{95+i} = \text{sgn}(T^i R_4), \quad i = 0,\dots,5, 
\end{equation}
whereas $c_{40} = -1$.
 The requirement that the logarithmic divergence cancels leads to a very simple identity for the UT integrals in four--dimensional kinematics, namely
\begin{equation}
    \tilde{I}_{33} = \sum_{j=0}^{5}c_{95+i} \tilde{I}_{27+i}.
    \label{eq:HexIdentity6D}
\end{equation}
Interestingly, comparing with the well--known identity \eqref{eq:HexIdentity4D}, we find after dividing \eqref{eq:HexIdentity6D} by~$\Delta_6$ 
\begin{align}
    \frac{\Delta_{5,j}}{\Delta_6} = \left(\frac{1}{2} \sum_{k=1}^6 (S^{-1})_{jk}\right)^2
\end{align}
on the momentum twistor parametrisation. Hence, after taking into account the signs $c_{i}$, the identity we find from the canonical differential equation for the UT integrals is equivalent to~\eqref{eq:HexIdentity4D}. 

At the level of the weight--four symbol of the hexagon integral, the four--dimensional limit implies
\begin{align}
    \lim_{D_{\text{ext}}\rightarrow 4}\mathcal{S}^{(4)}(\tilde{I}_{33}) = \frac{1}{2}\sum_{i=1}^3 T_i \left(\Omega \otimes W_{101}\right),
\end{align}
where $\Omega$ is the linear combination of weight--three symbols of box--integrals given in eq.~\eqref{eq:BoxCombination}.

\subsection{Dependencies between letters in four--dimensional kinematics}
In addition to the divergent letters, there are also some additional dependencies between letters that arise in four--dimensional kinematics. For simplicity, we only discuss the case in which all $c_i=-1$. All other cases can be reached by Galois transformations, that send the corresponding square roots to their negatives (and hence invert the respective odd letters). The full list of identities read

\begin{align}
    T^{i-1}\left(\frac{W_{101}}{W_{64} W_{65}} \right)&=1, \quad i=1,\dots,6\notag\\
    T^{i-1}\left( W_{89} W_{91} W_{93} \right)&= 1, \quad i=1,2 \notag\\
    T^{i-1}\left( \frac{W_{71} W_{73} W_{83}}{W_{60} W_{66} W_{77}} \right)&= 1, \quad i=1,\dots,6 \notag\\
   T^{i-1}\left( \frac{W_{77} W_{80} W_{101} W_{102} W_{103}
    }{W_{61} W_{64}W_{67} W_{70}W_{73} W_{76}} \right) &=1,\quad i=1,2,3.
\end{align}
On the support of these identities, the hexagon alphabet reduces to 89 independent letters.
\section{Discussion and outlook}
\label{sec:outlook}
In this paper we have investigated the one--loop on--shell hexagon integral in $D$--dimensional kinematics as functions of the nine independent Mandelstam variables using the method of canonical differential equations. To find a canonical basis, we have used dimension--shift identities and normalised the different integrals of the hexagon family by their leading singularities in order to construct a pure function basis of uniform transcendentality. To translate the canonical differential equations into dlog form, we have constructed the hexagon alphabet from the previously known one--mass five--point alphabet \cite{Abreu:2020jxa, Chicherin:2021dyp} and the by now well--understood recursive structure of one--loop alphabets \cite{Chen:2022fyw}.

By imposing the absense of non--physical singularities, we have then determined the boundary values for the hexagon family for a judiciously chosen point in the Euclidean region. This boundary information, together with the differential equation matrix in dlog form in principle allows us to evaluate the hexagon family at arbitrary points on any sheet of the Riemann surface at any order of $\epsilon$ in terms of Chen iterated integrals.

For a particular subsection of the Euclidean region, we provide a proof--of--concept implementation of the numerical evaluation. Using integration--by--parts, we have set up one--fold integral representations for the weight--three and the weight--four parts of the hexagon family, employing the analytically known weight--two solutions. We found perfect agreement with numerical evaluations obtained with \verb|AMFlow|~\cite{Liu:2022chg}.

Using the differential equation, we have also calculated the symbol of the hexagon integral at weight three, confirming the well-known result of Ref.~\cite{Dixon:2011ng}, and for the first time at weight four. 

Finally, we have studied the limit of four--dimensional external kinematics, for which the nine Mandelstam variables are no longer independent but satisfy a Gram determinant constraint. We show how the identity relating the four--dimensional hexagon integral to one--mass pentagon integrals comes out of the differential equation and discuss the additional relations between alphabet letters that emerge in the four--dimensional limit.

In conclusion, we find it very convenient to work in a $D$--dimensional setting and only fix the dimensionality to four at the very end by going to a kinematic point which satisfies the Gram determinant constraint. Even though the integral family is slightly bigger in $D$ dimensions, we profit from a more systematic understanding of the alphabet and manifest realisations of permutation invariance. Clearly, the next logical step will be to extend this analysis to the two--loop case. While at one loop, the single additional independent integral in $D$--dimensional kinematics is not a vast complication, at two--loops there will certainly be a trade--off between additional integrals and alphabet letters on the one hand and clearer structures and free movement in the nine--dimensional kinematic space on the other hand. Since the correct identity allowing us to remove the additional integrals in the four--dimensional limits can be derived from the knowledge of the differential equation, the $D$--dimensional analysis can be of use even if one is interested purely in four--dimensional results.  

For phenomenological applications of our results, the hexagon integral has to be evaluated on the physical $2\rightarrow4$ scattering sheet. While it is governed by the same differential equations as on the Euclidean sheet, it will be necessary to either determine the values of the entire integral family at a new boundary point in this physical region or to carry the information from our boundary point to the physical region via some appropriate analytic continuation. We leave this analysis for future work. 

Furthermore, having the full $D$--dimensional one-loop hexagon alphabet and its four--dimensional limit at our disposal, it will be very interesting to investigate them with respect to cluster algebra structures. While it has been appreciated for some time that cluster algebras play an important role for scattering amplitudes in $\mathcal{N}=4$ super--Yang--Mills theory \cite{Arkani-Hamed:2012zlh,Golden:2013xva} (see also Ref.~\cite{He:2022ctv}), the application to single Feynman integrals is more recent and still relies on a case--by--case analysis \cite{Chicherin:2020umh,He:2021eec}. 

On a similar note, it would be interesting whether the alphabet can be constructed efficiently from the Landau equations. It can be easily checked that the vanishing locus of any of our letters solves the Landau equation but it is not clear whether this can be used to set up an algorithm that systematically constructs these letters from the Landau equations only. At one loop, it appears that the minors of the Cayley determinant provide the necessary building blocks but it would be highly desirable to extend this construction to higher--loop cases.


\section*{Acknowledgments}

We thank Simon Badger, Wojcieh Flieger, Thomas Gehrmann, David Kosower, William J. Torres Bobadilla, Tiziano Peraro, Yingxuan Xu, Yang Zhang and Simone Zoia for useful discussions. This research received funding from the European Research Council (ERC) under the European Union's Horizon 2020 research and innovation programme (grant agreement No 725110), {\it Novel structures in scattering amplitudes}.

\bibliographystyle{JHEP}
\bibliography{refs}
\end{document}